\documentclass[12pt,aps,a4paper,preprint, showpacs, showkeys, superscriptaddress]{revtex4}
\usepackage{graphicx}
\usepackage[breaklinks=true,colorlinks=true,linkcolor=blue,urlcolor=blue,citecolor=blue]{hyperref}
\usepackage{amsmath}

\begin{document}

\title[Effect of AC-Stark shift in optical dipole trap  on absorption imaging...]{Effect of AC-Stark shift in optical dipole trap on absorption imaging of trapped atoms }

\author{K Bhardwaj}
\email{kavish@rrcat.gov.in}
\author{S P Ram}
\author{S Singh}
\affiliation{Laser Physics Applications Section, Raja Ramanna Centre for Advanced Technology, Indore, 452013, India}
\author{V B Tiwari}
\author{S R Mishra}
\affiliation{Laser Physics Applications Section, Raja Ramanna Centre for Advanced Technology, Indore, 452013, India}

\affiliation{Homi Bhabha National Institute, Training School Complex, Anushakti Nagar, Mumbai, 400094, India}

\begin{abstract}
In the present work, the effect of AC-Stark shift (i.e. light-shift) in optical dipole trap on in-situ absorption probe imaging of the trapped atoms has been investigated. We have calculated  the light-shift of various energy levels of $^{87}$Rb atoms relevant for trapping in an optical dipole trap (ODT). The spatial varying intensity of the ODT beam results in position dependent light-shift (i.e. AC-Stark shift). Such modifications in the energy levels of the atom result in modification in the absorption cross-section, which finally modifies the optical density (OD) in in-situ absorption imaging at a given wavelength of imaging probe beam. Incorporating the light-shift in analysis of images, the correct number of atoms in the optical dipole trap has been estimated for the experimentally observed absorption images of the trapped cloud. The in-situ imaging and this work can be useful in estimating the instantaneous loss rate from the trap during the evaporative cooling of trapped atom cloud.      

\end{abstract}


\vspace{2pc}

\maketitle

\section{Introduction}

Ultracold atoms are presently under active investigation for interesting physics and several technological applications in precision measurements \cite{Inguscio2013,Poli2011}, quantum information \cite{Bloch2012a}, etc. Among others, optical dipole traps (ODTs) are important tools for generation, trapping and manipulation of ultracold atoms \citep{Inguscio2013,Bloch2012a,Grimm2000a, Garraway2000,Kuppens2000}. As compared to magnetic traps, ODTs provide magnetic sub-level independent trapping \cite{Inguscio2013}, easier manipulation of potential and high trap frequencies, which are useful for faster evaporative cooling to achieve Bose-Einstein Condensation (BEC) degenaracy \cite{Arnold2011a,Barrett2001a,Chaudhuri2007,Landini2012,Yamashita2017}. The optical dipole traps with single potential well are used in preparation of ultracold degenerate atomic gases \cite{Grimm2000a} via trapping and evaporative cooling. A variant of the optical dipole trap (ODT) with multiple wells is optical lattice, which is used for atom trapping to study several interesting physics and condensed matter aspects \cite{Grimm2000a,Inguscio2013,Poli2011}.  An ODT is formed due to the interaction of a spatially varying intensity profile of trapping laser beam with laser field induced dipole moment of the atom. In an ODT, the frequency of trapping laser beam governs the structure of the atom trapping potential and dictates whether atoms will be trapped in the high intensity region (if frequency of trapping laser is red-detuned to the atomic transition) or in low intensity region (if frequency of trapping laser is blue-detuned to the atomic transition).  

Since atoms are trapped in high intensity region in ODTs with red-detuned trapping laser beam, the trapped atoms undergo a significant light-shift in energy levels. This is in contrast to the situation in magnetic traps as well as to the dipole traps with blue-detuned frequency, where atoms are trapped at minimum of the field. In an ODT with red-detuned trapping laser beam, the spatially varying intensity of trapping beam is expected to result in position dependent light-shift of the energy levels  \cite{Garraway2000}. Since light-shift of the levels not only governs the trapping potential of an ODT, but also modifies the transition frequencies of various transitions of the atom in the ODT, a precise knowledge of the shift of various energy levels due to AC-Stark effect in ODT is crucial for correct estimation of trapping potential as well as in analysis of absorption images of trapped cloud. 

During our experimental work on ODT of $^{87}$Rb atoms, the laser cooled atom cloud from a magneto-optical trap (MOT) was trapped in an ODT generated by focusing a red-detuned Gaussian laser beam of wavelength 1064 nm. In this ODT,  the number of atoms were obtained by probing the trapped cloud by in-situ absorption imaging technique. The number of atoms obtained from the analysis of images (without considering light-shift) was much smaller than one would expect in view of the number and temperature of atoms in MOT and knowing the ODT parameters. This stimulated us to consider the effect of light-shift (AC-Stark shift) due to dipole laser beam on in-situ absorption probe imaging and estimation of number of atoms trapped in ODT. This kind of study has not been reported earlier to the best of our knowledge. Therefore, the work presented here can be useful for many researcher who aim to perform in-situ absorption imaging in the ODT and/or aim to monitor instantaneous loss rate in the ODT during evaporative cooling.

In the work presented in this paper, a detailed calculation of light-shift of various energy levels of $^{87}$Rb atom due to interaction with an ODT of a  single focused laser beam at 1064 nm has been carried out in order to evaluate the dipole trap potential and the modified absorption cross-section at absorption probe laser wavelength. These results have been utilized in analysis of absorption probe images of atoms trapped in our ODT. After performing the image analysis with modified absorption cross-section, the correct number of atoms trapped in ODT in our experiments has been estimated. 

\section{Experimental}
\label{sec:expt}

\begin{figure}[h]
 \centering
 \includegraphics[scale=0.5]{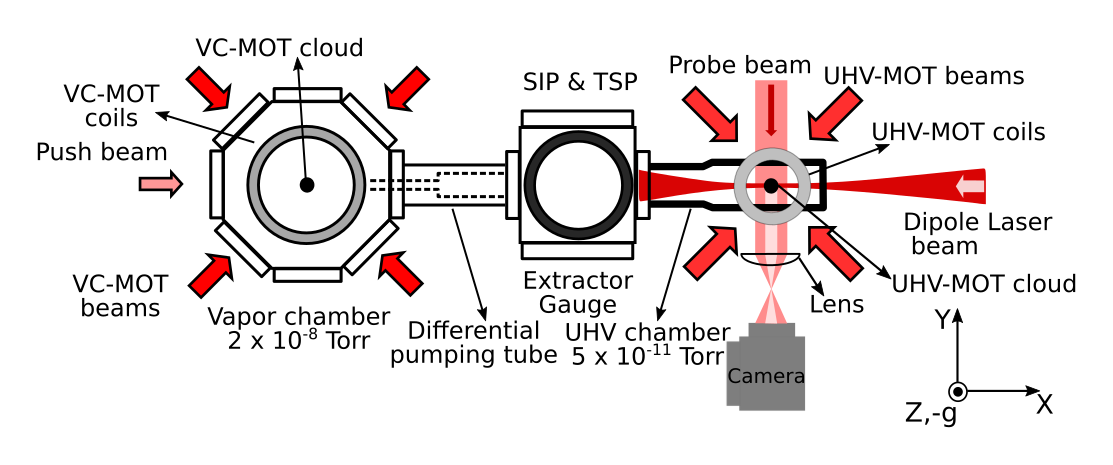}
 \caption{\label{fig:expt_setup} Schematic of the experimental setup. The absorption probe setup is aligned in a perpendicular direction to the dipole beam propagation direction. Complete optical layout is not shown to maintain clarity of the setup. SIP $\&$ TSP:  sputter ion pump and Titanium sublimation pump.}
\end{figure}

\begin{figure}[t]
 \centering
 \includegraphics[scale=0.15]{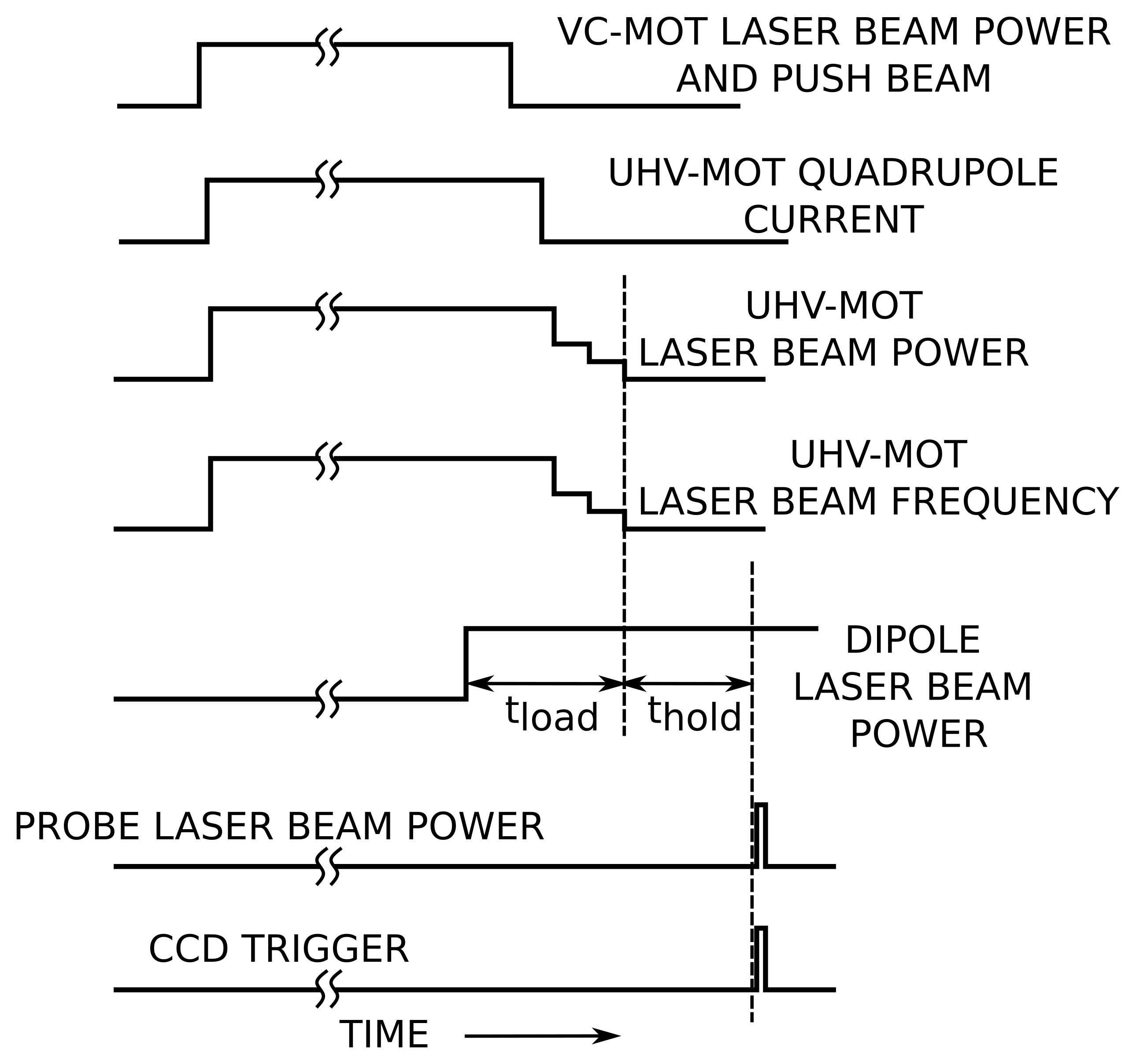}
 \caption{\label{fig:pulse_seq}The experimental timing sequence showing the various control signals used for loading and detection of the dipole trap.}
\end{figure}

The laser cooled atoms have been trapped in a single beam optical dipole trap by focusing a 1064 bm laser beam on a MOT cloud. An schematic of the experimental setup is shown in Fig.\ref{fig:expt_setup}. It is a double magneto-optical trap (MOT) setup consisting of two MOT chambers in horizontal plane, connected through a narrow tube (known as differential pumping tube). One chamber is octagonal shape made of stainless steel (at pressure $\sim$2$\times10^{-8}$ Torr with Rb vapor) and serves as a vapor chamber (VC) for source MOT (called VC-MOT), whereas the other chamber is a glass cell (at pressure $\sim$5$\times$10$^{-11}$ Torr) which serves as UHV chamber for collection MOT (called UHV-MOT) and ODT purpose. The differential pumping tube is used between two chambers to maintain the pressure difference. The laser beams for formation of MOTs in both the chambers are derived from output of two tapered amplifier systems (BoosTA, Toptica Germany). These amplifiers are seeded by two low power extended cavity diode lasers (ECDL) (DL 100, Toptica, Germany). The frequency of one of the seed lasers is locked using Doppler free saturated absorption spectroscopy (SAS) technique at red side of the cooling transition (5S$_{1/2}$ F=2$\rightarrow$P$_{3/2}$ F$^\prime$=3) with detuning of 2.5$\Gamma$ ($\Gamma = 2\pi\times$ 6 MHz, natural line width of $^{87}$Rb). The amplified output of this laser from first amplifier serves the purpose of cooling laser. The other seed laser is operated at resonance frequency of repumping transition (5S$_{1/2}$ F=2$\rightarrow$P$_{3/2}$ F$^\prime$ = 2), whose amplified output from the second amplifier serves as repumping laser. The outputs from both the amplifiers are mixed and then split into two beams. These two beams are expanded and further split to generate desired VC-MOT beams and UHV-MOT beams. Several acousto-optic modulators (AOMs) are used in the laser beam path to control the frequency and power of the laser beams. For VC-MOT formation three beams are used in retro-reflection geometry, whereas, for UHV-MOT, six independent beams are used. Another ECDL system operated at ~1 GHz red detuned from cooling transition (5S$_{1/2}$ F=2$\rightarrow$P$_{3/2}$ F$^\prime$=3) is used as pushing laser to transfer the atoms from  VC-MOT to UHV-MOT. Two pairs of current carrying coils, VC-MOT coil and UHV-MOT coil, are used to create required quadrupole magnetic field distributions for VC-MOT and UHV-MOT. The Rb vapor is generated in the octagonal chamber by passing a current of 3.2 A through the Rb dispensers connected to a vacuum feed-through \cite{Rapol2001}. By locking the lasers to the desired frequencies, flowing suitable currents through the MOT coils, and aligning pushing beam properly, we obtain $\sim 1 \times 10^{8}$ number of atoms in the UHV-MOT at a temperature of $\sim$ 200 $\mu$K. The temperature of the cloud is reduced further by allowing the cloud to evolve in optical molasses.

\begin{figure}[b]
  \centering
    \includegraphics[scale=0.3]{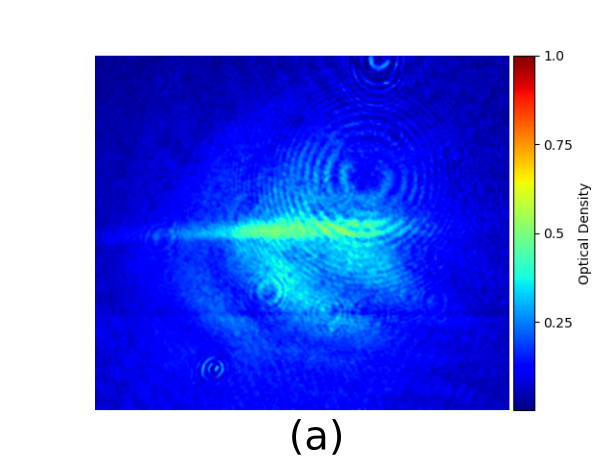}
    \includegraphics[scale = 0.3]{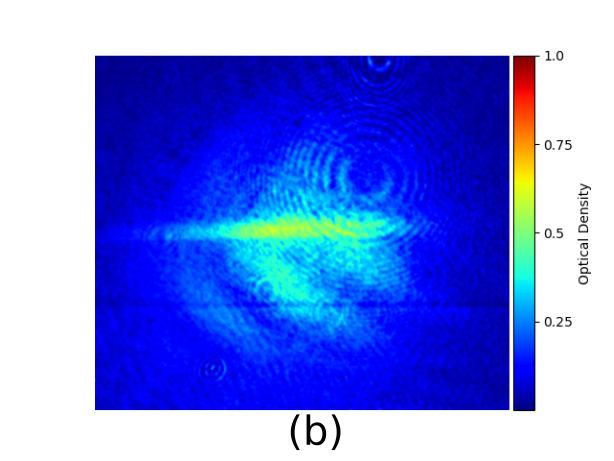}
    \includegraphics[scale=0.3]{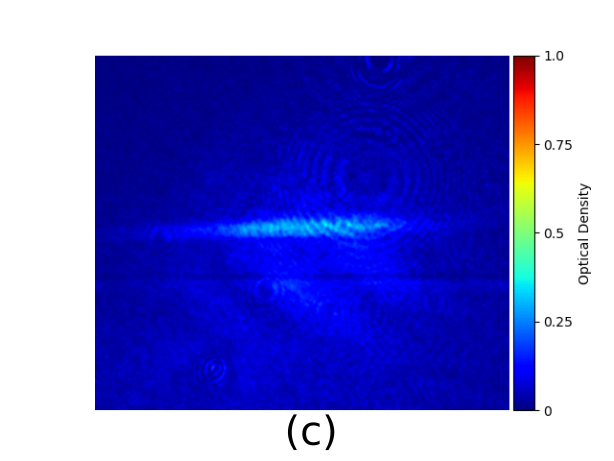}

  \caption{\label{fig:expt_cloud} The optical density images generated from the absorption probe detection setup for dipole beam power of  (a)19.7 W,  (b) 24.9 W and (c) 27.7 W.}
\end{figure}

A fiber laser (Model :YLR-1064-100-LP,IPG Photonics, USA) has been used to generate the dipole trap beam \cite{Grimm2000a,Hung2015}. The laser output beam is expanded three times and then focused using a lens of 200 mm focal length. The lens is kept on a manual micrometer controlled translation stage having three degrees of freedom, so that the focal position of the dipole beam can be tuned across the UHV-MOT. The focal spot size (1/e$^2$ radius) of the ODT was ~17 $\mu$m with a trap depth of 0.336 mK/W. The focal spot size was measured using a beam profiler (M/s OPHIR Optronics). 

A part of the cooling laser beam is used as absorption probe beam by appropriately passing through an AOM and expanding suitably. This beam is aligned in a direction perpendicular to the direction of propagation of dipole laser beam (as shown in Fig. \ref{fig:expt_setup}) to image the atom cloud  in dipole trap. The process and control sequence \cite{Sofikitis2011,Szczepkowicz2009,Grimm2000a,Kuppens2000,Tanasanchai2018} for loading the dipole trap is shown in Fig. \ref{fig:pulse_seq}. The automation of the ODT loading was controlled by a field programmable gate array (FPGA) based controller system in conjunction with an industrial personal computer. All the processes such as VC-MOT loading, transfer of atoms to UHV-MOT, MOT compression, optical molasses stage, loading atoms in ODT, and finally detection of number of atoms trapped in ODT are executed through this controller. The  switching ON/OFF of the dipole laser beam was done with a mechanical shutter, which was controlled by a suitable pulse generated by a function generator.

As discussed earlier \cite{Kuppens2000}, area of the equipotential surface represented by U $\sim$ k$_{B}$T$_{MOT}$ should govern the loading rate in ODT, where $T_{MOT}$ is the temperature of atom cloud in the MOT. The area in the ellipsoidal surface is larger at positions away from the focal point of the ODT beam. Therefore, in our experiment, we have displaced the focal point of the ODT beam for the best loading of the ODT. We have noted that $\sim$ 8 mm shift of the focal position is appropriate to maximize the ODT loading at 27.7 W trapping beam power.

The optical dipole trap (ODT) was loaded by switching-on the dipole laser beam on MOT cloud. The trapped cloud in ODT was imaged by absorption probe imaging technique. A fine straight line \cite{Grimm2000a} at the centre of the image of atom cloud was observed at dipole laser power of $\sim$ 10 W (corresponding to a trap depth of $\sim$ 3.36 mK). This showed the trapping of cold atoms in ODT. 

As the power of ODT beam was increased, the optical density (OD) in the absorption images increased initially, but started decreasing as power was further increased (Fig. \ref{fig:expt_cloud}). After the analysis of images, it was found that at higher ODT beam power the number of trapped atoms in ODT (estimated without considering AC-Stark effect) did not follow the expected trend in trapped number vs power of ODT beam. To our surprise, the OD (and estimated number) in images ( e.g. Fig. \ref{fig:expt_cloud}(b) and \ref{fig:expt_cloud}(c)) decreased with increase in power of ODT beam. Therefore, we improved our analysis of absorption images by incorporating AC-Stark shift (light-shift) of energy levels of Rb-atoms due to ODT beam. At higher dipole trap beam power, it is expected that light-shift can modify the absorption of probe laser significantly and lead to reduction in OD in the images of trapped atom cloud. Hence, to find the accurate OD in absorption probe imaging, an exact calculation of the AC-Stark shift and effective absorption cross section is necessary. This has been discussed in forthcoming section.    

\section{Results and discussions}
Due to considerably high intensity in ODT beam, the AC-Stark shift (light-shift) in atomic states due to ODT beam can be significant which may modify the absorption cross-section of atoms at probe beam wavelength during the in-situ absorption probe imaging of cloud in ODT. Thus it is important to calculate the AC-Stark shift (light-shift) in atomic energy states and use it to calculate the probe absorption cross-section and finally estimate the number of atoms in the ODT accurately. 
\begin{table}
\centering
\caption{\label{tbl:lightshift}Most dominant transitions for calculation of peak light-shift values for F =2 and $m_F$ = 0 level due to dipole laser beam wavelength at 1064 nm, power 1 W and focused to 1/e$^2$ spot size of 17 $\mu$m.}
   \includegraphics[scale=0.35]{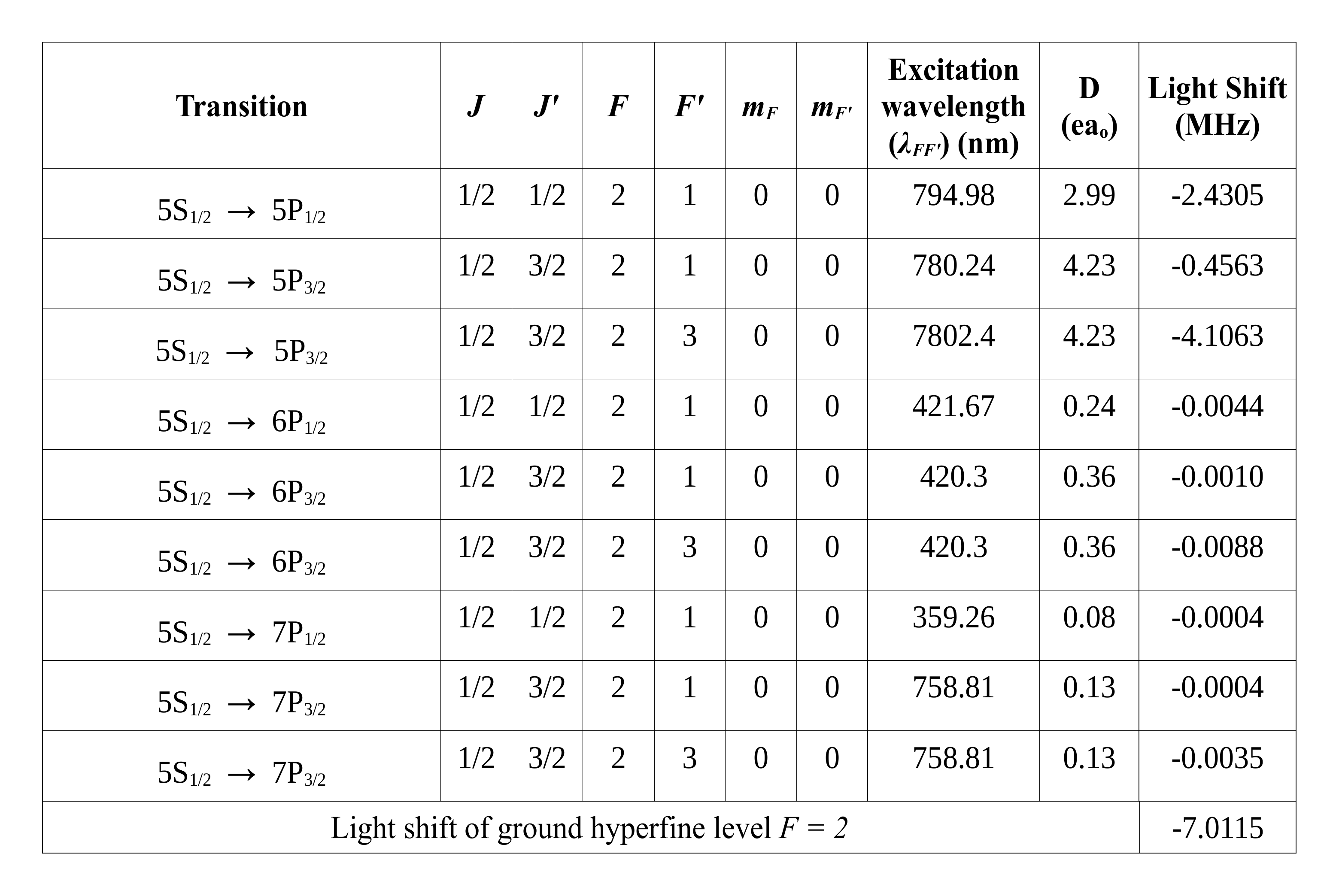}
   
\end{table}

\subsection{AC-Stark shift}
 If $I(x, y, z)$ is the spatial intensity of the laser used for dipole trap, the light-shift of any hyperfine level (F, $m_F$) is given by\cite{Shih2013,Steck2012},

\begin{multline}
    \Delta_{Fm_F}(x,y,z) = \frac{3\pi I(x,y,z)c^2}{2}\sum_{F^\prime m_{F^\prime}}\frac{\alpha\beta}{\omega_{FF^\prime}}(2F+1)(2F^\prime+1)(2J^\prime+1)\\
        \begin{pmatrix} F^\prime & 1 & F\\m_{F^\prime} & 0 & m_{F} \end{pmatrix}^2_{3j}
    \begin{Bmatrix} J & J^\prime & 1\\ F^\prime & F & 3/2 \end{Bmatrix}^2_{6j}
\end{multline}

\begin{equation}
    \alpha = \left(\frac{1}{\omega_{FF^\prime}+\omega}+\frac{1}{\omega_{FF^\prime}-\omega}\right)  ;  \beta = \frac{2J+1}{2J^\prime+1}\frac{2.02613\times10^{18}}{\lambda_{FF^\prime}^3}D^2
\end{equation}
 
where, $\omega_{FF^\prime}$ is the angular frequency of transition between (F,$m_F$) and ($F^\prime, m_F^\prime$) levels, $D$ is the electric dipole moment (in atomic units), $\lambda_{FF^\prime}$ is the excitation wavelength for transition, $\omega$ is the dipole laser angular frequency, $\left(\right)$3j and $\left\{\right\}$6j are Wigner symbols and c is the speed of light in vacuum. The summation is carried out over all the dipole allowed transitions from (F,$m_F$) state to various ($F^\prime, m_F^\prime$) states.

\begin{figure}
    \centering
    \includegraphics[scale=0.4]{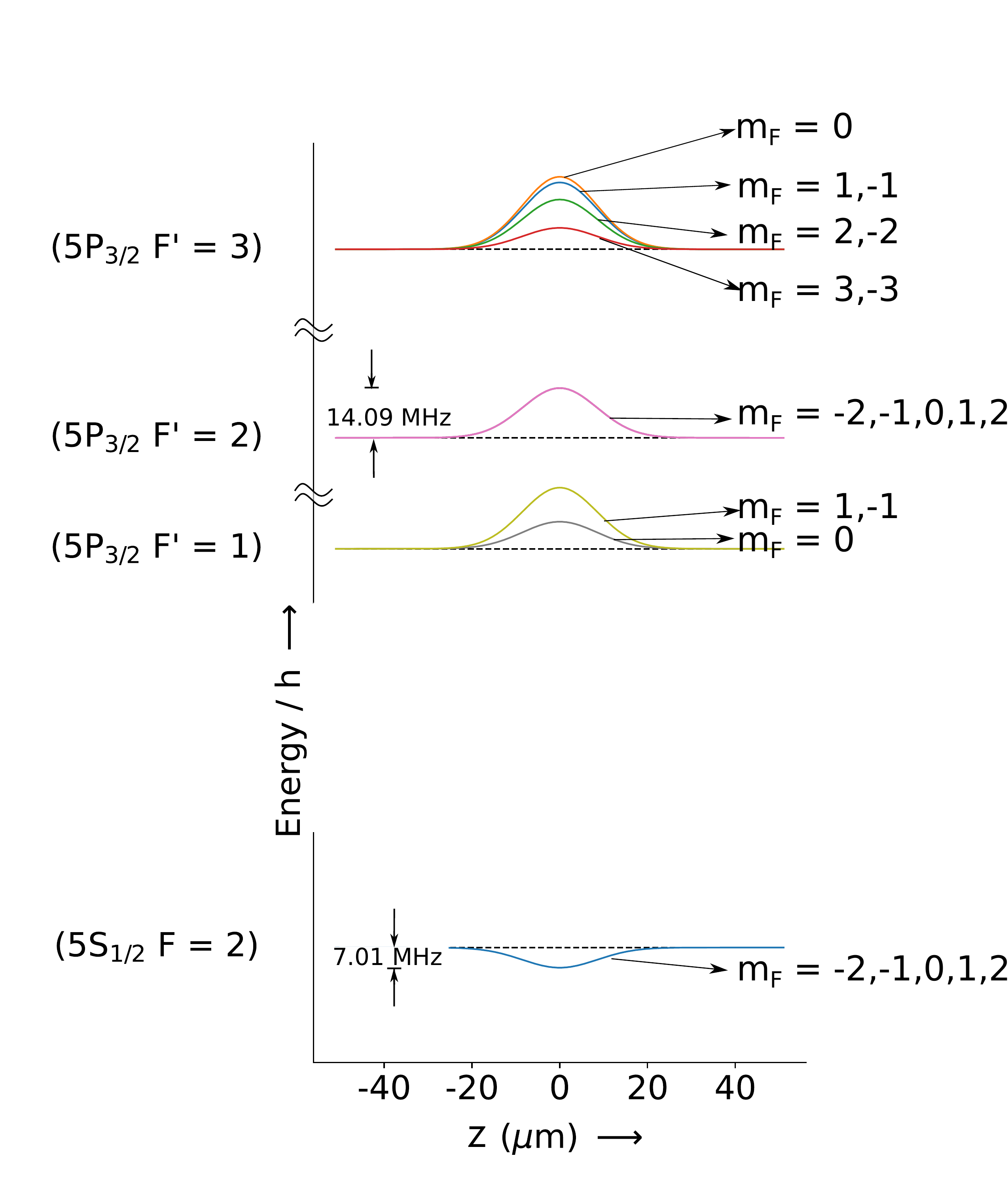}
    \caption{\label{fig:light_shift} The variation in AC- Stark shifts of various levels of $^{87}$Rb atom in the radial direction at the focus of the dipole trap beam at beam power of 1 W and spot-size of (1/e$^{2 }$ radius) 17 $\mu$m. The dashed lines show energy levels without dipole laser beam.}
\end{figure}

Various transitions have been accounted for calculation of light-shift for a
particular hyperfine level. Table \ref{tbl:lightshift} shows the transitions used for calculation of
the peak light-shift for the ground state hyperfine level (5S$_{1/2}$ F=2,
h$_{F}$ = 0) due to a dipole trapping laser beam at 1064 nm wavelength, 1 W power and
focused to 1/e$^{2}$ spot size of 17 $\mu$m. The peak AC- Stark shift of ground hyperfine state (5S$_{1/2}$ F=2) is -7.01 MHz at focus, which results in a trap depth of $\sim$ 0.336 mK (8.4 mK) for 1 W (24.9 W) laser power.

Due to the spatial intensity profile of the dipole laser beam, light-shift of
levels is position dependent and its variation is different along the dipole beam propagation
direction (\textit{x}-axis) as compared to the transverse direction (\textit{y or z}-axis). The variation in light-shift (AC-Stark shift) with position \textit{z} for different states (for ground state and few excited states) is
shown in Fig.\ref{fig:light_shift} .

Since, atomic polarizability for light atom interaction is \textit{m$_{F
}$} dependent, the calculated light-shift is also \textit{m$_{F}$} dependent.
The light-shifts are also dependent on geometry of dipole laser beam with respect to quantization axis \cite{Safronova2011}. In these calculations, trap beam polarization is linearly polarized along the quantization axis and beam propagation is
perpendicular to quantization axis. In our calculations, we obtained that
different \textit{m$_{F}$}$_{ }$levels in ground states shift equally, but shift in the
excited states (5P$_{3/2 }$ \textit{F' = 1}) and (5P$_{3/2 }$ \textit{F' = 3}) with
different \textit{m$_{F}$} sublevels is different. Since the light-shift in ground and excited states modifies the transition wavelengths
considerably, its consequence on absorption cross-section of atoms in the dipole trap at absorption probe wavelength is significant.

\begin{figure}
    \centering
    \includegraphics[scale = 0.45]{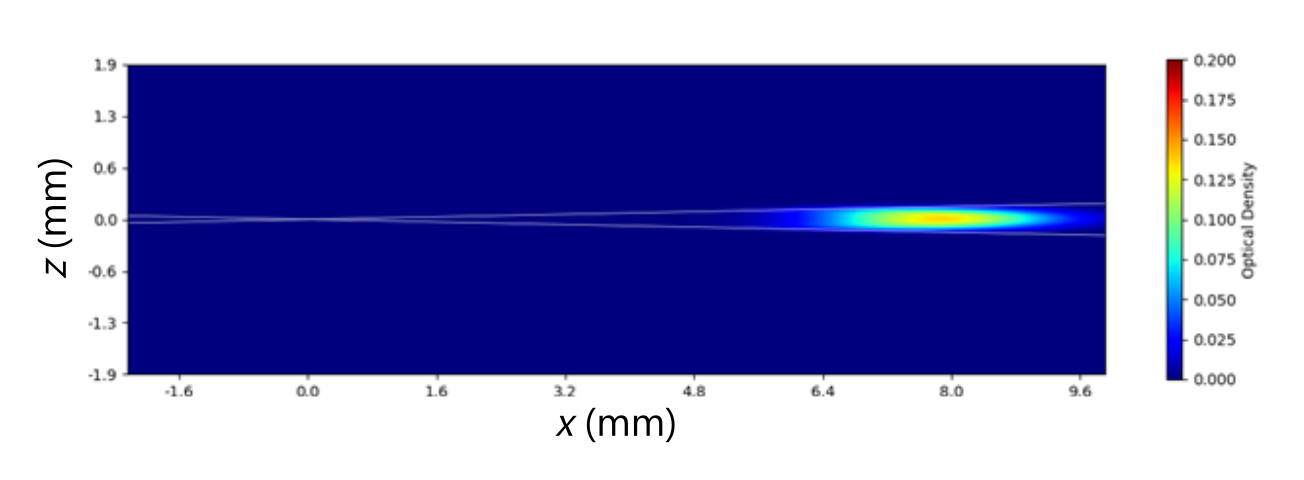}
    \caption{\label{fig:displaced} Optical density image generated by considering position dependent AC-Stark shift and effective absorption cross section due to a dipole laser beam of
power 27.7 W and 1/e$^{2 }$ radius 17 $\mu$m. The white line shows the expansion of 1/e$^{2}$ radius
of the dipole beam. The OD image shows the atom cloud is displacement by 8 mm from the focal position of the ODT beam.}
\end{figure}

\subsection{Effective absorption cross-section and optical density}

The trapped atom cloud in ODT was in-situ imaged by absorption probe imaging technique using a probe laser resonant to unperturbed 5S$_{1/2}$ F=2 to 5P$_{3/2}$ F'=3 transition. If we ignore the effect of AC-Stark shift on Rb energy levels, the absorption cross-section ${\sigma{}}_o$ in absorption probe imaging is taken to be constant. The measured optical density OD(x,z) (for probe beam propagating along y-axis) is used to find the number of trapped atoms using the relation N = (${\iint OD(x, z)dxdz}$)/${\sigma_0}$. This method to find the number of atoms (N) in the ODT needs modification if AC-Stark shift of energy levels is considered. The AC-Stark shift of energy levels is position and intensity dependent which results in position and intensity dependent effective absorption cross-section ${\sigma{}}_{eff}$(x,y,z). It is given as  

\begin{equation}
\sigma_{eff}(x,y,z) = \sum_{F^\prime m_F}\frac{\sigma_{F^\prime m_{F^\prime}}}{\left(1+4\left(\frac{\omega_{probe}-(\omega_{F^\prime m_F}-S_{F m_F}(x,y,z)+S_{F^\prime m_F}(x,y,z))}{\Gamma}\right)^2\right)}.
\end{equation}

where ${\sigma{}}_{eff}$(x,y,z) is the effective absorption cross-section for the transition $(F, m_F)$ to $(F^\prime, m_F^\prime)$. Here summation is carried out over all dipole allowed transitions from (F=2,m$_{F}$) states to (F$^{'}$ =1,2,3 , m$_{F '}$) states. The parameter ${\sigma{}}_{F^{'}m_{F^{'}}}$ is the absorption cross-section for transition from (F, m$_{F}$) state to (F$^{'}$, m$_{F '}$) state. These absorption cross sections for Rb atom transitions were calculated using standard formulae and were also verified using ratios of  strengths for various transitions\cite{Metcalf2003}.

Using the effective absorption cress-section ${\sigma{}}_{eff}$(x,y,z), the optical density (OD) images can be generated theoretically using the the equation

\begin{equation}
    OD(x,z) = \int \sigma_{eff}(x,y,z) n(x,y,z) dy .
    \label{eq:OD_eff}
\end{equation}

Here y-axis is the direction of propagation of absorption probe laser beam, $n(x, y, z)(= n_0 e^{{-\frac{x^2}{2\sigma_x^2}} {-\frac{y^2}{2\sigma_y^2}}{-\frac{z^2}{2\sigma_z^2}}})$ is the density of the cloud with $n_0$ as peak density and $\sigma_i$ is r.m.s width in $i$ (x, y, z) direction. The integration is carried over y-direction (propagation direction of the absorption probe (Fig. \ref{fig:expt_setup})). $n_0$ can be obtained from experimentally measured peak optical density. 

\subsection{Optical dipole trap potential and its surface}

\begin{figure}[h]
 \centering
 \includegraphics[scale=0.45]{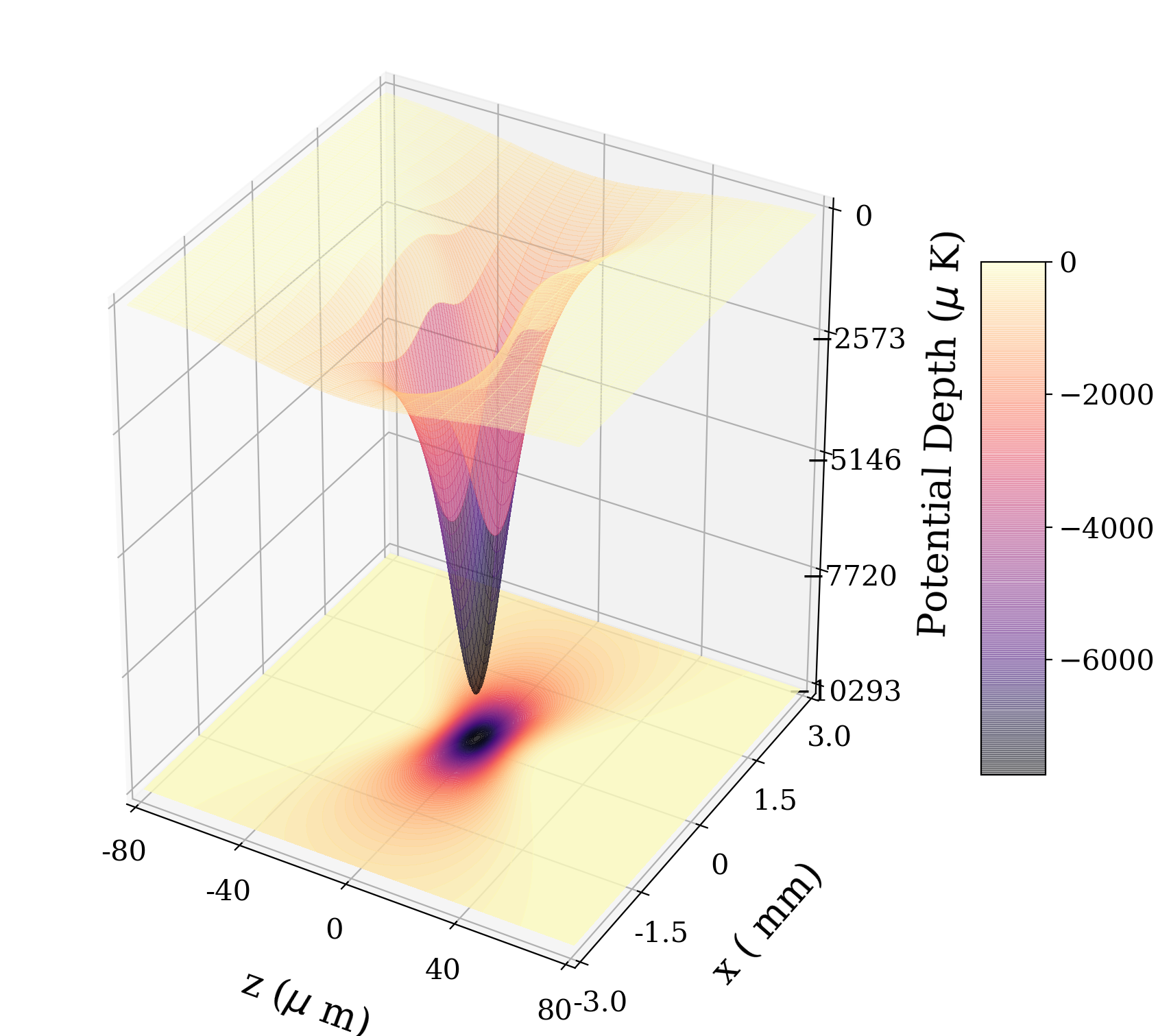}
 \caption{\label{potential}Three dimensional view of the trapping potential for the ground state ($5S_{1/2}$ F = 2, $m_{F}$) due to dipole laser beam wavelength at 1064 nm for power $\sim$ 25 W and 1/e$^2$ spot size of 17 $\mu$m.}
\end{figure}

The trapping potential in ODT follows the light-shift of the ground state. The trap depths along radial and propagation direction of the focused trapping beam for the ground state (5S$_{1/2}$ F = 2, $m_F$) are shown in Fig. \ref{potential}.

\begin{figure}[b]
    \centering
    \includegraphics[scale=0.3]{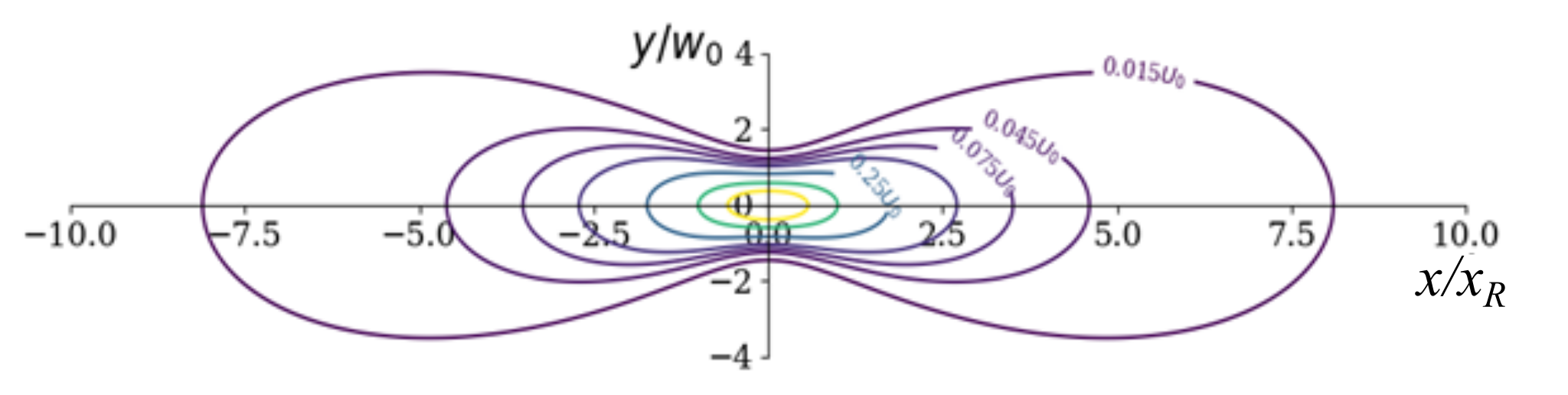}
\caption{\label{fig:equipotential} A cross-section of the equipotential surface of the ODT potential formed by the focused Gaussian beam (optical dipole beam propagating in $x$ direction having Rayleigh range as $x_R$). }
\end{figure}

The trapping potential contour can be visualized in Fig.\ref{fig:equipotential}. It has been pointed out earlier \cite{Szczepkowicz2009} that the potential surface is an ellipsoidal surface with larger area away from focal position. Kuppens et al \cite{Kuppens2000} suggested that initial loading rate of ODT is proportional to the effective surface area of ODT. In view of this, it can be beneficial for ODT loading to align dipole trap beam on the MOT such that MOT position was slightly shifted from focal position of the ODT beam\cite{Szczepkowicz2009}. Figure \ref{fig:displaced} shows such a simulated OD image of the cloud in which cloud is shown shifted from focal point of ODT beam.

\subsection{Number of trapped atoms vs power of dipole laser beam}

\begin{figure}[h]
    \centering
    \includegraphics[scale=0.4]{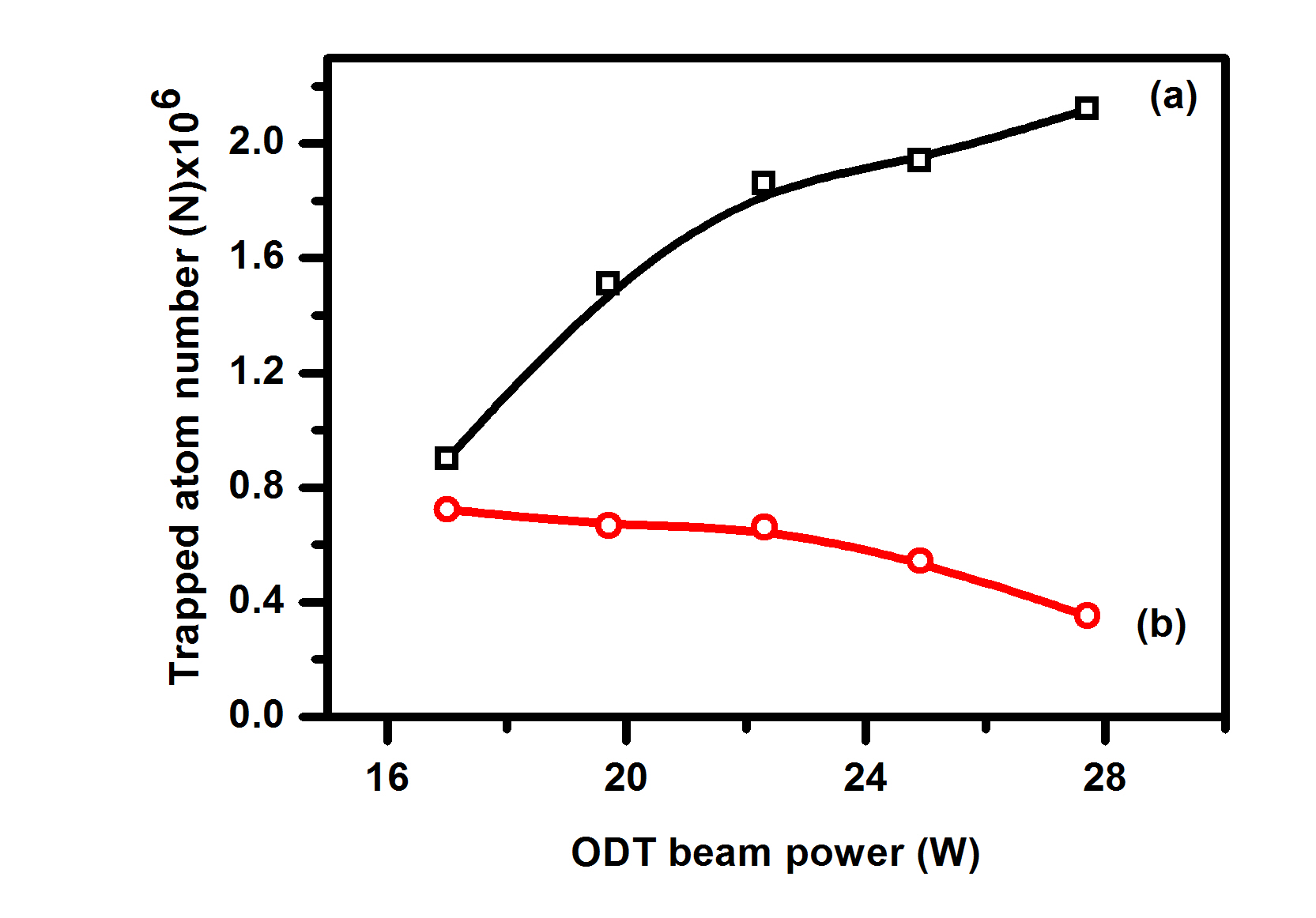}
\caption{\label{fig:N_vs_power} Variation in the number of atoms trapped in optical dipole trap with the power of the ODT beam, (a): after incorporating AC-Stark shift and (b): without incorporating the AC-Stark shift.}
\end{figure}

In order to correctly estimate the number atoms (N) trapped in ODT, the effective absorption cross-section ${\sigma{}}_{eff}$ calculated after knowing position and intensity dependent AC-Stark shift has been used in the analysis of experimentally observed absorption images of the trapped atom cloud. The peak number density $n_0$ was obtained by keeping theoretically calculated OD (from eqn.(\ref{eq:OD_eff})) equal to the experimentally observed OD. The number of atoms in the trap was then estimated using expression N $= n_0 (2\pi)^{3/2}\sigma_x \sigma_y \sigma_z$.

The estimated number of atoms with and without incorporating AC-Stark shift are shown in Fig. \ref{fig:N_vs_power} for different values of power in ODT beam. It is evident that there is considerable difference in number of atoms estimated with and without considering light-shift. For example, at 27.7 W power, the number of atoms trapped in ODT are found to be 2.12 $\times10^{6}$ with AC-Stark shift included in the analysis, whereas the number of atoms estimated without taking AC-Stark shift into account is 3.5 $\times10^{5}$. The difference between these two numbers is significant. Hence, inclusion of AC-Stark shift in analysis is important.

\section{Conclusion}

In conclusion, a detailed calculation of AC-Stark shift (i.e. light-shift) of different hyperfine levels of $^{87}$Rb atom has been performed. After knowing the AC-Stark shift in the ground state, the potential landscape for trapping the cold $^{87}$Rb atom in a single beam ODT has been obtained accurately. As the light-shift in ground and excited states modifies the transition wavelengths considerably, the effective absorption cross-section for the probe laser beam has been evaluated. Finally, this effective absorption cross-section has been used to estimate the correct number of atoms in the atom cloud trapped in the ODT. The difference in the number of atoms estimated with and without incorporating light-shift shows that it is essential to consider AC-Stark shifts for correct estimation of number of trapped atoms in dipole trap.

We thank V. Bhanage, P. P. Deshpande, S. Tiwari, L. Jain, and A. Pathak (all from RRCAT) for developing the controller system used in the experiments and Vivek Singh for useful technical discussions.



\end{document}